\documentclass[11pt,a4paper]{article}
\pdfoutput=1
\usepackage{jheppub}

\usepackage{latexsym}
\usepackage{amsmath, mathrsfs}
\usepackage{graphicx, slashed}

\title{It is a Graviton! or maybe not}
\author{Ricky Fok$^1$, Caroline Guimar\~aes$^1$, Randy Lewis$^1$ and Ver\'onica Sanz$^{1,2}$}
\affiliation{$^1$Department of Physics and Astronomy, York University, Toronto, ON, Canada, M3J 1P3}
\affiliation{$^2$Theory Division, Physics Department, CERN, 1211 Geneva 23, Switzerland}

\abstract{The discovery of Kaluza-Klein (KK) gravitons is a smoking gun of extra dimensions. Other scenarios, however, could give rise to spin-two resonances of a new strongly-coupled sector and   act as impostors. In this paper we prove that a spin-two resonance does not couple to the Standard Model through dimension-four operators.  We then show that the massive graviton and its impostor both couple to the Standard Model through the same dimension-five operators. Therefore the spin determination is identical. Nevertheless, we also show that one can use the ratio of branching ratios to photons and to jets for distinguishing between KK gravitons and their impostors. The capacity to distinguish between KK gravitons and impostors is a manifestation of the breakdown of the duality between AdS and strongly-coupled theories.}

\newcommand{\be}{\begin{equation}}
\newcommand{\ee}{\end{equation}}
\newcommand{\bea}{\begin{eqnarray}}
\newcommand{\eea}{\end{eqnarray}}

\begin{document}

\maketitle
\section{Introduction}

What is the smoking gun of extra dimensions at the Large Hadron Collider (LHC)? The obvious answer is that extra dimensions at the TeV scale would indicate gravity is strong near that scale, and black hole formation or other effects of strong gravity would be possible. That would be spectacular and convincing evidence for new dimensions of space-time. Alas, the setting of strong gravity is plagued with uncertainties and the LHC may not be able to access it \cite{LisaMeade}.  Instead, the smoking gun for extra dimensions would be the discovery of Kaluza-Klein (KK) excitations of fields propagating in the bulk of the extra dimension. Models vary in terms of bulk content, but the common feature of all models is that the graviton propagates in the extra dimension and, hence, KK gravitons\footnote{Interest for spin-two resonances goes beyond uncovering extra-dimensions, as they can contribute to the $t \bar{t}$ forward-backward asymmetry \cite{Grinstein:2012pn}.} are the signature of new dimensions of space-time. 

What are the properties of KK gravitons that one can look for? KK gravitons, which we denote here by $G$, are massive spin-two particles, and one could use the angular distributions of the KK graviton decay products to determine the spin~\cite{Allanach:2000nr}. One can also look at selection rules related to the spin structure of the resonance~\cite{LisaWise}.   

But other new physics could be behind massive spin-two states. For example, a new strongly-coupled sector could produce the analogue of the $f_2$ meson in QCD~\cite{Nakamura:2010zzi}, a resonance with $J^{PC}=2^{++}$ just like a KK graviton. Therefore, spin determination is insufficient to claim the discovery of new dimensions. Also, it has been speculated \cite{Foadi:2008xj} that a spin-2 resonance arising from technicolor theories could be misidentified as a KK graviton. The graviton $G$ and the impostor $\hat{G}$ would both be massive spin-two resonances. On the other hand, $G$ couples to the stress tensor of  the Standard Model (SM) and, at first sight, one would think that the impostor $\hat{G}$ could exhibit a broader range of couplings. But, as we show in this paper, that is not the case. 

In this paper we prove:

\begin{itemize}
  \item Once Lorentz and SM gauge symmetries are imposed,  $\hat{G}$ cannot couple through a dimension-four term in the Lagrangian.
  
  \item  If one further assumes that the composite sector respects the flavor and CP symmetries of the SM, then $G$ and $\hat{G}$ couple to the {\it same} operators of the SM at leading order, namely the same dimension-five operators.
  
  \item Although the operators coupling to $G$ and $\hat{G}$ are the same, the coefficients are not. In the case of the KK graviton $G$, the coefficients are given by the Planck mass and the overlap of wavefunctions in the extra dimension. For the $\hat{G}$ case, those coefficients are largely unknown, related to the UV structure of the strongly-coupled theory responsible of the appearance of $\hat{G}$.  Nevertheless, we find a robust prediction for a ratio of coefficients in the $G$ case, and this ratio is a {\it real} smoking-gun for extra dimensions. 
\end{itemize} 

Besides helping to determine whether a new dimension has been discovered, our result has interesting consequences for the holographic approach to technicolor, or lack thereof.

The paper is structured as follows. In Sec.~\ref{prop}, we show that $G$ and $\hat{G}$ have the same propagation, and in Secs.~\ref{Gco} and~\ref{Ghatco} we describe their couplings to the SM particles. We then show ways of distinguishing between them in Sec.~\ref{dist}, and describe aspects of the holographic picture between $G$ and $\hat{G}$. 

\section{The propagation of $G$ and its impostor}\label{prop}

In this section we show that the KK graviton $G$ and the spin-two meson from new strong interactions $\hat{G}$ have the same propagation, described by the Fierz-Pauli Lagrangian for massive spin-two resonances~\cite{Fierz-Pauli}. In the  Fierz-Pauli Lagrangian, a spin-two field is described by a rank-two symmetric and traceless tensor,
\bea
\hat{G}_{\mu\nu}=\hat{G}_{\nu\mu} \, \ , \, \hat{G}_{\mu}^{\mu} =0 \ . \label{traceless}
\eea
Moreover, the following condition must be satisfied for $\hat{G}_{\mu\nu}$ to have positive energy
\bea
\partial_{\mu} \hat{G}^{\mu\nu} = 0 \ . \label{divergence}
\eea
The propagation of $G$ is identical. The argument is as follows. Let us assume for simplicity that there is a new extra-dimension, denoted by $z$, which is compactified in an interval $z \in (z_{UV},z_{IR})$. We will sometimes denote those limits as the UV(IR) {\it brane}, as one can localize fields on those four-dimensional (4D) manifolds. We then define the set of five-dimensional (5D) factorizable metrics,
\bea
ds^2=w^2(z) \, (\eta_{\mu \nu} d x^{\mu} d x^{\nu}-dz^2) \ ,  \label{geom}
\eea 
where $w(z)=1$ or $z_{UV}/z$ respectively for a flat or AdS extra dimension. In general, $w(z)$ is a constant or decreasing function of $z$. 

Since the graviton field in an extra-dimensional theory has a massless zero mode (the 4D graviton), the 5D graviton field has Neumann boundary conditions on both sides of the interval.  Kaluza-Klein dynamics is obtained by studying fluctuations around the Minkowski metric in Eq.~(\ref{geom}),
\bea
\eta_{\mu \nu} \to \eta_{\mu \nu} + h_{\mu \nu}(x,z) \ .
\eea
The equation of motion of the graviton field is given by the Einstein equation, 
\bea
G_{MN} = 0
\eea 
where $M,N=0,1,2,3$ and $5$. Note that the metric is conformally flat, which allows the following separation
\bea
G_{MN} = G^{flat}_{MN} + \delta G_{MN}[\nabla w(z), \nabla \nabla w(z)],
\eea 
where $G^{flat}_{MN}$ is the Einstein tensor in Minkowski space-time, and it contains the Fierz-Pauli equation for the graviton in flat space-time. Now $ \delta G_{MN}$ contains only covariant derivatives of the warp factor $w(z)$. Because the warp factor is only a function of the extra dimension coordinate $z$, only derivatives with respect to $z$ will appear in  $ \delta G_{MN}$. Then, upon KK decomposition of the graviton field, $G_{\mu \nu}(x,z) = \sum_n G^{n}_{\mu \nu}(x)\chi_n(z)$, terms in  $ \delta G_{MN}$ appear in the differential equation for the 5D wavefunction $\chi_n(z)$ of excited KK gravitons \cite{KKgravRS}, while the kinetic term in four dimensions remains the same as the flat space-time case, i.e. the Fierz-Pauli equation. Therefore, all KK excitations behave as 4D Fierz-Pauli fields, and the same equations as Eqs.~(\ref{traceless}) and~(\ref{divergence}) apply to $G$\footnote{We would like to note that this result has been obtained in the flat space case \cite{Han:1998sg,Giudice:1998ck} as well as the Randall-Sundrum case~\cite{KKgravRS}.}. Finally, note that we could easily generalize this argument to $D>5$.


\section{The coupling of $G$ to the Standard Model} \label{Gco}

In this section we describe the couplings of the KK graviton to matter. Those couplings are in general model-dependent functions of the geometry of the extra dimension and localization of fields in the bulk of the extra dimension. Nevertheless, one can extract general aspects of those couplings, as we discuss below.

The graviton couples to matter through the energy stress tensor. The Lagrangian describing the interactions is

\begin{equation}
\mathscr{L}_{int}=-\frac{c_{i}}{M_{eff}} G^{\mu \nu} T^{i}_{\mu \nu}, \label{LG}
\end{equation}
where $T^{i}_{\mu \nu}$ is the 4D stress tensor of SM species $i=b$, $f$, $H$ (gauge bosons, fermions, scalars). $M_{eff}$ is the effective Planck mass suppressing the interactions, and we are going to focus on the case
\bea
M_{eff}  \gtrsim m_{G} \simeq \textrm{ TeV} \ ,
\eea
and assume $M_{eff}/m_G$ is at least 2 or 3, indicating that the effective theory has a range of validity beyond the first resonance $G$.
Finally, the $c_i$ are functions of the overlap of the $G$ resonance with the SM fields in the bulk of the extra dimension. 
 
The relevant $G$-SM-SM interaction terms can be found in 
\bea
T^{f}_{\mu \nu} &\supset& \frac{i}{2} \, \bar{\psi} \gamma_{\mu}\partial_{\nu}\psi + (\mu \leftrightarrow \nu),\\ 
T^{A}_{\mu \nu} &\supset& -F_{\mu}^{ \;\rho} F_{\rho \nu}, \\ 
T^{H}_{\mu \nu} &\supset& \partial_{\mu} H \partial_{\nu} H + (\mu \leftrightarrow \nu) \ .
\label{Tmunu}
\eea
Note that in $T_{\mu\nu}$ there are also terms with more than two fields as well as terms proportional to electroweak symmetry breaking (EWSB), i.e. proportional to $m_{W,Z}$.

What about the values of the coefficients $c_{i}$? Assuming the extra-dimensional geometry can be expressed in the general form of Eq.~(\ref{geom}), one can estimate the coefficients as follows\cite{Gherghetta:2010cj}:
\begin{enumerate}
  \item {\bf Brane fields: } If the SM field lives on a brane located at $z_*$
  \bea
  c \simeq w(z_*)/w(z_{IR})
  \eea
  where $z_*$ is the location of the brane, $z_*=z_{IR, UV}$. In flat extra dimensions, $w=1$ and there is no parametric suppression on either brane. In warped extra dimensions, $w(z_{IR}) \ll$ $w(z_{UV})$ and 
  \bea
 \frac{w(z_{UV})}{w(z_{IR})} \simeq \frac{M_{Pl}}{M_{eff}}  \simeq \frac{M_{Pl}}{\textrm{  TeV}}  \ .
  \eea
  \item {\bf Bulk fields in flat extra-dimensions:} In flat extra dimensions, Kaluza-Klein number is conserved as long as there are no localized boundary terms. In that case, if the SM field lives in the bulk of a flat extra dimension, then the coupling $G$-SM-SM vanishes,
  \bea
  c = 0 \textrm{ with KK conservation.} 
  \eea
  On the other hand, without KK conservation, the overlap of fields in the extra dimension would be of order one, leading to $c \simeq 1$. 
  \item {\bf Bulk fields in warped extra-dimensions: } If now some fields live in the bulk of extra dimensions, their coupling to $G$ depends on their localization or de-localization in the bulk. Note that $G$ is localized near the IR brane at $z_{IR}$.
\begin{itemize}
  \item Coupling to IR-localized fields: $c \simeq 1$.
  \item Coupling to massless gauge bosons: suppressed by the effective volume of the extra dimension~\cite{gap-metrics}, 
  \bea
  c \simeq \frac{1}{\int_{z_{UV}}^{z_{IR}} w(z) dz} \ .
  \eea
 In AdS, the suppression is $\log(\frac{z_{IR}}{z_{UV}})\simeq\log(\frac{M_{Pl}}{M_{eff}})$. In flat space, the suppression is the entire volume of the extra dimension.
  \item Coupling to UV-localized fields: suppression of order 
 \bea
c \simeq \left(\frac{z_{UV}}{z_{IR}}\right)^a=\left(\frac{TeV}{M_{Pl}}\right)^a \ ,
 \eea
 where $a>1$.
For example, in Randall-Sundrum, the coupling of $G$ to UV-localized fermions is given by
\bea
c_{f} \propto \epsilon^{2 |\nu-1/2|}
\eea
where $\nu<$ -1/2 for UV-localized fermion zero modes and $\epsilon\simeq TeV/M_{Pl}$. Similarly for UV localized scalars with bulk mass parameter $\nu<1$, 
\bea
c_{\phi} \propto  \epsilon^{-2(1-\nu)}
\eea
where $\nu=z_{UV} \, M_{\psi,\phi}$, where $M_{\psi,\phi}$ is the bulk fermion (scalar) mass. 
\end{itemize}
\end{enumerate}

\section{The couplings of the impostor $\hat{G}$ to the Standard Model}\label{Ghatco}

In the previous section, we discuss which operators couple to the resonance $G$, and how the coefficients of these operators strongly depend on how the SM particles are localized in the bulk of the extra dimension, or localized on one of the boundaries. For example, in warped extra dimensions, only  fields with some support near the IR brane at $z_{IR}$ would have sizable overlap with the KK resonance. That includes fields on or near the IR brane, and delocalized fields (i.e. fields with a flat profile in the extra dimension).
 
The impostor $\hat{G}$ is a resonance of a new sector which confines near the electroweak scale, at $M_{conf}$. As we want to discuss the role of $\hat{G}$ as an impostor of $G$,  we identify $M_{conf}$ with $M_{eff}$. 

In principle, one could imagine $\hat{G}$ coupling to SM particles in a very different fashion than $G$, since it is not constrained by the form of interaction in Eq.~(\ref{LG}). But, as we discuss in this section, Lorentz and gauge invariance determine the couplings of $\hat{G}$ to be dimension-five operators, and if one further assumes flavor and CP invariance, $\hat{G}$  couples to the same operators contained in Eq.~(\ref{LG}).

\begin{table}[t]
\centering
\begin{tabular}{|c|c|c|}
\hline
    $\hat{O}^{decay}_{\mu \nu}$   & CP & coefficients \\ \hline
$i\bar\psi \gamma_{\mu}\partial_{\nu}\psi $ & $+$ & $ \hat{c}_f^+$\\
$i\bar\psi \gamma^5 \gamma_{\mu}\partial_{\nu}\psi  $& $-$& $\hat{c}_f^-$\\
$F_{\mu}^{ \;\rho} F_{\rho \nu} $    & $+$ & $ \hat{c}_A^+$\\
 $ \epsilon_{\alpha \beta \mu \delta} F_{\nu}^{\; \delta} F^{\alpha \beta} $ & $-$ &$ \hat{c}_A^-$\\
 $\partial_{\mu} H \, \partial_{\nu} H$ & + &  $\hat{c}_{\phi}$\\  \hline
\end{tabular}
\caption{Flavor-invariant operators up to dimension 5 that could lead to two-body $\hat{G}$ decays. If we further assume the composite sector preserves CP invariance, the remaining operators are the same structures contained in the stress tensor.}
\label{table:f2fermions}
\end{table}

After imposing  Lorentz and gauge  invariance, $\hat{G}$ exhibits no interactions with fermions, gauge bosons and scalars at the level of dimension four operators. For example, operators such as $\bar{\Psi} \gamma_{\mu} \gamma_{\nu} \Psi$  or $F_{\mu\nu}$ (for abelian gauge groups) vanish due to properties in Eqs.~(\ref{traceless}). Also, interactions where the derivative acts on $\hat{G}$ vanish because of the conditions in Eq.~(\ref{divergence}).

Table \ref{table:f2fermions} shows all operators that could lead to two-body decays of $\hat{G}$ up to dimension 5 with no flavor violation. Up to CP conservation, the remaining operators are identical to those listed in Eqs.~(\ref{Tmunu}). 

It could be that the new physics responsible for $\hat{G}$ includes new sources of CP violation. In particular, a non-zero coefficient for the operator $\hat{c}_f^-$ in Table~\ref{table:f2fermions} would be  constrained by precision measurements of, for example, the kaon system. But those operators contain derivatives of the fermion, and by integrating out the massive resonance we would obtain a CP-violating four-fermion operator involving light quarks,
\bea
\sim \frac{\hat{c}_i^c \hat{c}_j^c}{M_{eff}^2} \, \frac{\hat{s}}{m_{\hat{G}}^2} \, \bar\psi^i \gamma_{\mu} \gamma_5 \psi^i  \bar\psi^j \gamma^{\mu} \gamma_5 \psi^j  \ ,
\eea
which is suppressed by $\hat{s}/m_{\hat{G}}^2$.
We would obtain a bound on the coefficient of the CP-violating operator~\cite{Gilad} 
\bea
c \lesssim 10^{-2} \, \frac{ M_{eff} \, m_{\hat{G}}}{\textrm{TeV}^2} 
\eea 
where we estimated $\sqrt{\hat{s}}\sim$ ${\cal O}$(GeV). 

The focus of this paper is the distinction between a KK graviton and its impostor so from now on we are going to assume that CP is an approximate, or exact, symmetry of the strong sector, and therefore the coupling of a $J^{CP}=2^{++}$ resonance to a CP-violating operator is suppressed.

\section{Distinguishing between the graviton and the impostor}\label{dist}

In the last two sections, we showed that $G$ and $\hat{G}$ couple to the same dimension-five operators.
What about dimension-six operators or higher? Obviously, they are suppressed by an extra power of the TeV scale, and their effect is sub-leading. Still, we could classify all dimension-six operators compatible with Lorentz and gauge invariance. Unfortunately, we do not know  the behavior of these operators, neither for gravity nor for a strongly-coupled theory. On the gravity side, those operators would arise as a consequence of quantum gravity loops, and their coefficients are therefore hard to estimate. On the strong-coupling side, one faces similar ambiguities. So, dimension-six operators cannot be a way to distinguish between $G$ and $\hat{G}$, and we need to look closely at dimension-five operators to find ways of disentangling signatures for extra dimensions from composite strongly-coupled dynamics.

\subsection{The ratio of decay to photons and gluons}

Although the form of the gravitational interaction is fixed, the coefficients of operators that couple to the KK graviton are model-dependent. To distinguish between $G$ and an impostor, we would need a model-independent prediction. In this section we show that such an observable is possible.
 
We define the following ratio:

\bea
R_{g/\gamma}=\frac{Br(\to g g)}{Br(\to \gamma \gamma)} = \frac{8 c_{g}^2}{c_{\gamma}^2} \ .
\eea

In extra dimensions $R$ is fixed to be 8, whereas for $\hat{G}$ there is no constraint on $R$. The argument is as follows: For any geometry in the form of Eq.~(\ref{geom}), the KK decomposition for spin-one particles leads to an equation of motion for the wavefunction  of the $n$th KK mode, $f_n(z)$~\cite{me-hol},
\bea
\partial_z \left(  w(z) \partial_z f_n(z)\right) = -m_n^2 w(z) f_n(z) \ .
\eea
If the spin-one field has a massless zero-mode, i.e. the 4D gauge symmetry is preserved by the compactification, then
\bea
m_0=0 ~~\to~~ w(z) \partial_z f_0(z) = \textrm{constant}
\eea
Once we take into account the boundary conditions, which are Neumann on both branes,  
\bea
\partial_z f_0(z) |_{UV} = \partial_z f_0(z) |_{IR} = 0 \ ,
\eea
there is only one solution
\bea
f_0(z)=C
\eea
where the constant $C$ is determined by imposing the canonical normalization for the 4D gauge field. 
In Randall-Sundrum models, the value of $c_{\gamma,g}$ is
\bea
c_{\gamma, g}=2 \frac{1-J_0(x_G)}{\log(z_{IR}/z_{UV}) x_G^2 |J_2(x_G)|}
\eea
where $x_G=3.83$ and $M_{eff}=M_{Pl} w(z_{IR})$. Here we see explicitly the suppression with $1/\int w d z$ mentioned in Sec.~\ref{Gco}. 

The discussion above is a consequence of QCD and electromagnetism in the bulk. On the other hand, one could imagine localizing electromagnetism and strong interactions on a brane. Electromagnetism is part of the electroweak group. Therefore, the photon, if stuck on a brane, should be stuck on the brane responsible for electroweak symmetry breaking, i.e. $z_{IR}$. That leads to numerous problems with compositeness effects showing up at the TeV scale and altering precision measurements. Still, this is the scenario that is searched experimentally for Randall-Sundrum models~\cite{Allanach:2000nr}. Also in this case, the ratio $R$ is 8.  

Finally, a situation where the gluon is stuck on the UV brane, whereas the photon is on the IR brane is phenomenologically ruled out since quarks are charged under both and would need a non-negligible overlap with both branes, only possible if de-localized.  

In summary, in any phenomenologically viable model we would have a prediction for this ratio, $R=8$. Let us now discuss some aspects of measuring this ratio.

In principle, $G$ could have a non-negligible branching ratio to light quarks. Gluons and quarks are seen as jets in colliders, and we would need to distinguish those to evaluate $R$. 

In the most successful Randall-Sundrum scenarios~\cite{BulkRS}, light quarks are UV-localized fields having a very small overlap with $G$. Hence s-channel production of $G$ is through gluons and the branching ratio to jets is to gluon-jets. 

\begin{figure}[h!]
\centering
\includegraphics[scale=0.3]{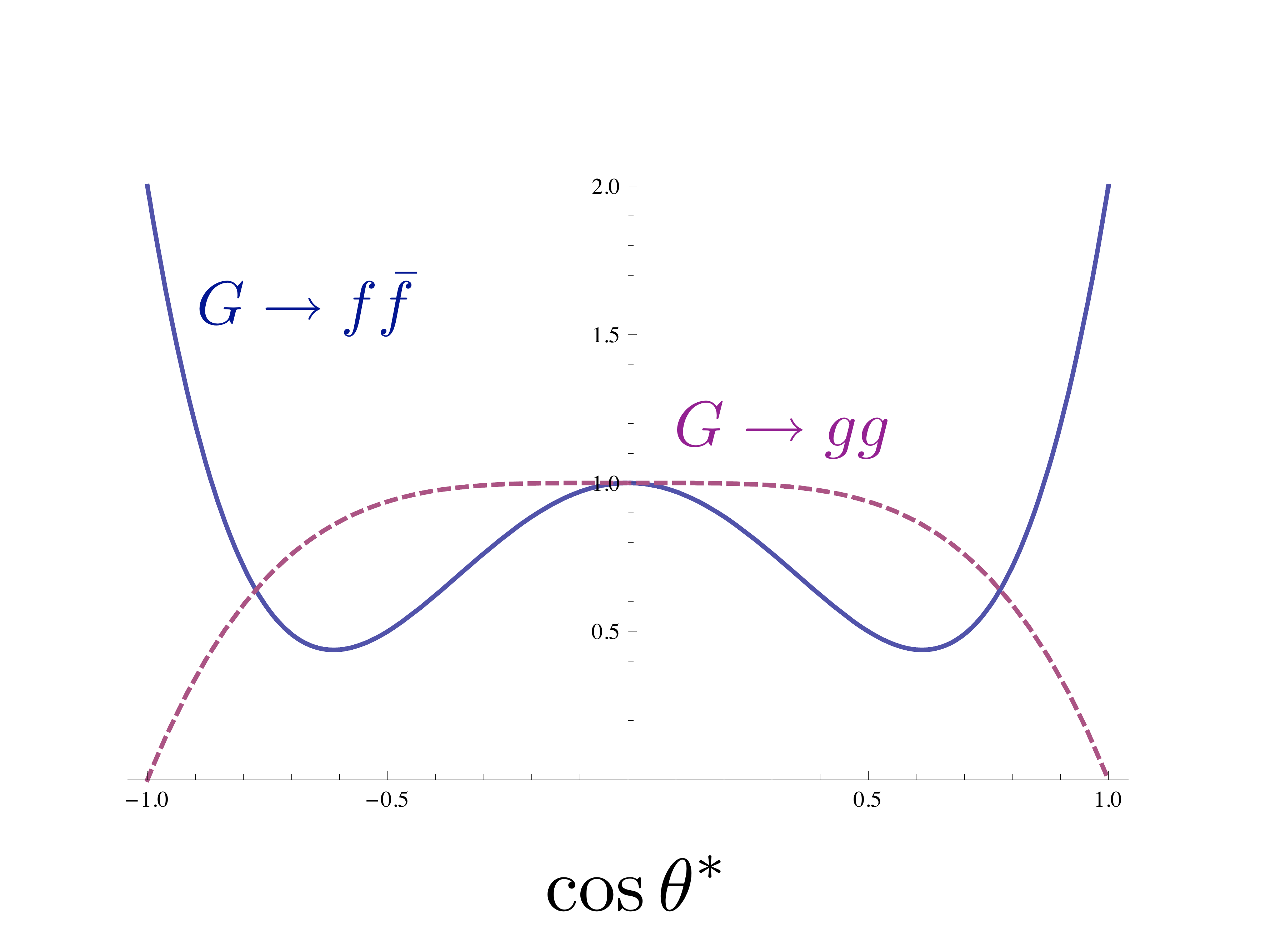}
\caption{The angular distributions for fermion  (solid line) and gluon (dashed line) final states.}
\label{distr}
\end{figure}

The assumption that $G$ has small couplings to the light generations is related to flavor issues and fermion mass generation. Nevertheless, there are scenarios where the light fermions have sizable couplings to KK physics and the flavor problem is solved by a choice of symmetries (see for example Ref.~\cite{Michele}). With significant couplings to light quarks and gluons, a heavy resonance would preferentially be produced  in quark-initiated processes, and one would have to disentangle the quark and gluon components of a dijet final state. One has two ways to attack this problem. First, the di-quark and di-gluon angular distributions are different 
\bea
\frac{d \sigma}{d \cos \theta^* } (q \bar{q} \to G \to f \bar{f}) = 1+\cos^2\theta^* \left(1-4 \sin^2\theta^*\right)
\eea 
\bea
\frac{d \sigma}{d \cos \theta^* } (q \bar{q} \to G \to gg) = 1-\cos^4\theta^*
\eea
where $\theta^*$ is the angle in the center of mass between the outgoing particle and the incident parton. In Fig.~\ref{distr} we show the two theoretical distributions.

Second, one could try to tag the jet as a gluon or quark using the techniques in Ref.~\cite{Matt-jet} which do not rely on angular distributions.

An early measurement of the spin relies on a sizable branching ratio of $G$ or $\hat{G}$ to photons~\cite{LisaWise}.  See also \cite{Hitoshi}. In the context of warped extra dimensions, one usually expects the third-generation quarks to be localized near $z_{IR}$. In these scenarios
\bea
\frac{Br(G\to t \bar{t})}{Br(G\to \gamma\gamma)} \propto \left(\int \frac{w(z)}{z_{UV}} d z \right)^2 \ ,
\eea
where the volume factor is $\cal O$(10's). Therefore, the dominant decay mode for $G$ would be to $t \bar{t}$, and one would need large luminosities for measuring both the spin and $R_{g/\gamma}$.

Finally, we would like to mention the typical production cross section for $G$. There is no model-independent prediction but a rather popular choice of extra-dimensional models is the implementation of Randall-Sundrum models in Madgraph~\cite{MG-RS}. With that choice of parameters, a 1 TeV resonance would be produced with a cross section of 2 pb for the LHC at 8 TeV. 

\subsection{Other spin-two states}

Strong interactions would produce a rich spectrum of resonances as we observe in QCD. In this section we discuss other spin-two resonances, both as a motivation to look for them, and as an illustration of the richness of dynamical electroweak symmetry breaking. 

In Randall-Sundrum models,
spin-2 resonances are the excitations of a graviton with quantum
numbers $J^{PC}=2^{++}$.  On the other hand, a QCD-like theory would
contain many spin-2 resonances including some with negative parity and/or
negative charge conjugation.  In QCD the lightest spin-2
resonances are $2^{++}$ and the next-lightest are $2^{-+}$.  All of these
QCD states are
readily understood from a simple quark model based on the Schrodinger
equation\cite{GodfreyIsgur}.  For up and down quarks, the $2^{++}$ states are
P-wave mesons
(an isosinglet named $f_2(1270)$ and an isotriplet named $a_2(1320)$)
while the $2^{-+}$ states are D-wave mesons
(an isosinglet named $\eta_2(1645)$ and an isotriplet named $\pi_2(1670)$).
Observation of a $2^{-+}$ resonance or charged $2^{++}$ resonances having a mass of order the electroweak
scale would be a clear indication of physics beyond a KK-graviton.

In QCD, those resonances would decay  predominantly into $f_2(1270)\pi$ or $3 \pi$. In the analogy of QCD with technicolor, decays to pions are decays to longitudinal $Z_L$ and $W_L$. Hence, those resonances produce a three-body decay and would not appear in the s-channel.  Now, a precise prediction for the decay rate is not possible
without knowledge of the underlying strong dynamics, but some general
insight can be obtained from
a rudimentary calculation of the $\pi_2-f_2$ mass difference\footnote{In QCD, the $a_2-f_2$ splitting is due to fine structure and it is of order 4\%.}.

\begin{figure}
\centering
\includegraphics[scale=0.4]{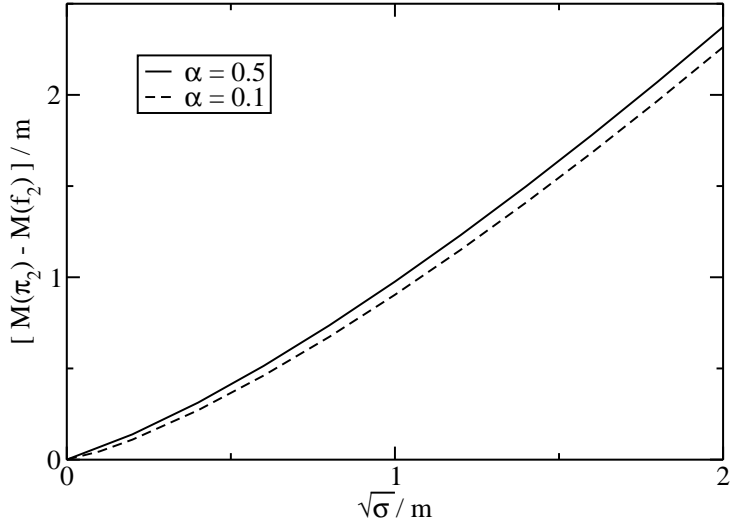}
\caption{The $\pi_2-f_2$ mass difference as a function of the square root of
the string tension, for two choices of the gauge coupling.
Both axes are in units of the constituent fermion mass.
At $\sigma=0$ the mass splitting does not vanish, but the tiny Coulomb
splitting is not visible in this graph.}
\label{fig:pi2f2}
\end{figure}

A naive rescaling of QCD to the electroweak scale (i.e.\ multiplying all masses
by $246{\rm GeV}/f_\pi ~\approx~2600$)
leads to a technicolor theory \cite{technicolor} that is opposed by
experimental data \cite{Chivukula:2012ug}, but other
strongly-interacting theories remain as viable options
\cite{Andersen:2011yj,Lewis:2011zb}, such as
the possibility of a walking or near-conformal theory \cite{walkingTC}.
Consider the Cornell potential \cite{Eichten:1979ms}

\[
V(r) = -\frac{4\alpha}{3r} + \sigma r
\]
where $\alpha$ represents the gauge coupling for the new strong interaction
and $\sigma$ is the string tension.
For a conformal theory, the string tension $\sigma$ must vanish.
(The current mass of the fermion must also vanish for a conformal theory,
but it is the nonzero constituent mass that appears in the Schrodinger equation.)
In QCD, the $f_2(1270)$ and $\pi_2(1670)$ have a $q\bar q$ separation that
is large enough to be dominated by the linear term (see figure 12 of
\cite{GodfreyIsgur}), but in a nearly-conformal theory where $\sigma$
is smaller those mesons would be in the Coulomb regime and the $\pi_2-f_2$
mass splitting would shrink.  To see this explicitly, we have used the
Mathematica code from \cite{Lucha:1998xc}
to solve the Schrodinger equation for a range of string tensions with
two different values for the coupling.  Numerical results for the mass
splitting are given in Fig.~\ref{fig:pi2f2} in units of the constituent fermion mass.
(In typical models, the constituent mass is 2 or 3 times the electroweak scale.)

For QCD, $\sqrt{\sigma}$ is between $m$ and $2m$ for any standard definition of
the constituent mass, and as a consequence the $\pi_2-f_2$ mass splitting is of
a comparable
size.  In the limit of vanishing string tension, the mass splitting in
Fig.~\ref{fig:pi2f2} becomes the Coulomb splitting, $M(\pi_2)-M(f_2)=\frac{5}{81}m\alpha^2$,
which is almost zero on the scale of the graph.
Because the string tension is so dominant,
measurement of the $\pi_2-f_2$ mass difference would provide valuable
information about the degree of conformality in the new strongly-interacting
theory.

\subsection{The holographic interpretation}

Models in warped extra dimensions are often used as an {\it analogue computer} for strong interactions. This duality between 4D strongly-coupled theories and 5D weakly-coupled theories with gravity was inspired by the AdS/CFT correspondence, but took hold on a more qualitative basis~\cite{Lisa-Porratti} and has been used to build models of QCD~\cite{AdSQCD,me-hol}, technicolor~\cite{HTC}, composite Higgs~\cite{composite-Higgs}, and even condensed matter systems~\cite{condensed-matter}, with some success.

What is the dual of a theory with gravity in extra dimensions? If the metric $w(z)$ is AdS, the dual is a 4D conformal theory, and compactification is the dual of spontaneous breaking of conformality, leading to a theory with massive resonances. If the metric is not the one of AdS, the 5D spacetime does not have the same isometries as the conformal group in 4D, and one expects no conformal behavior of the confining theory. Still, compactification of the extra dimension would be the dual description of confinement, and the appearance of massive resonances. 

\begin{figure}[h!]
\centering
\includegraphics[scale=0.2]{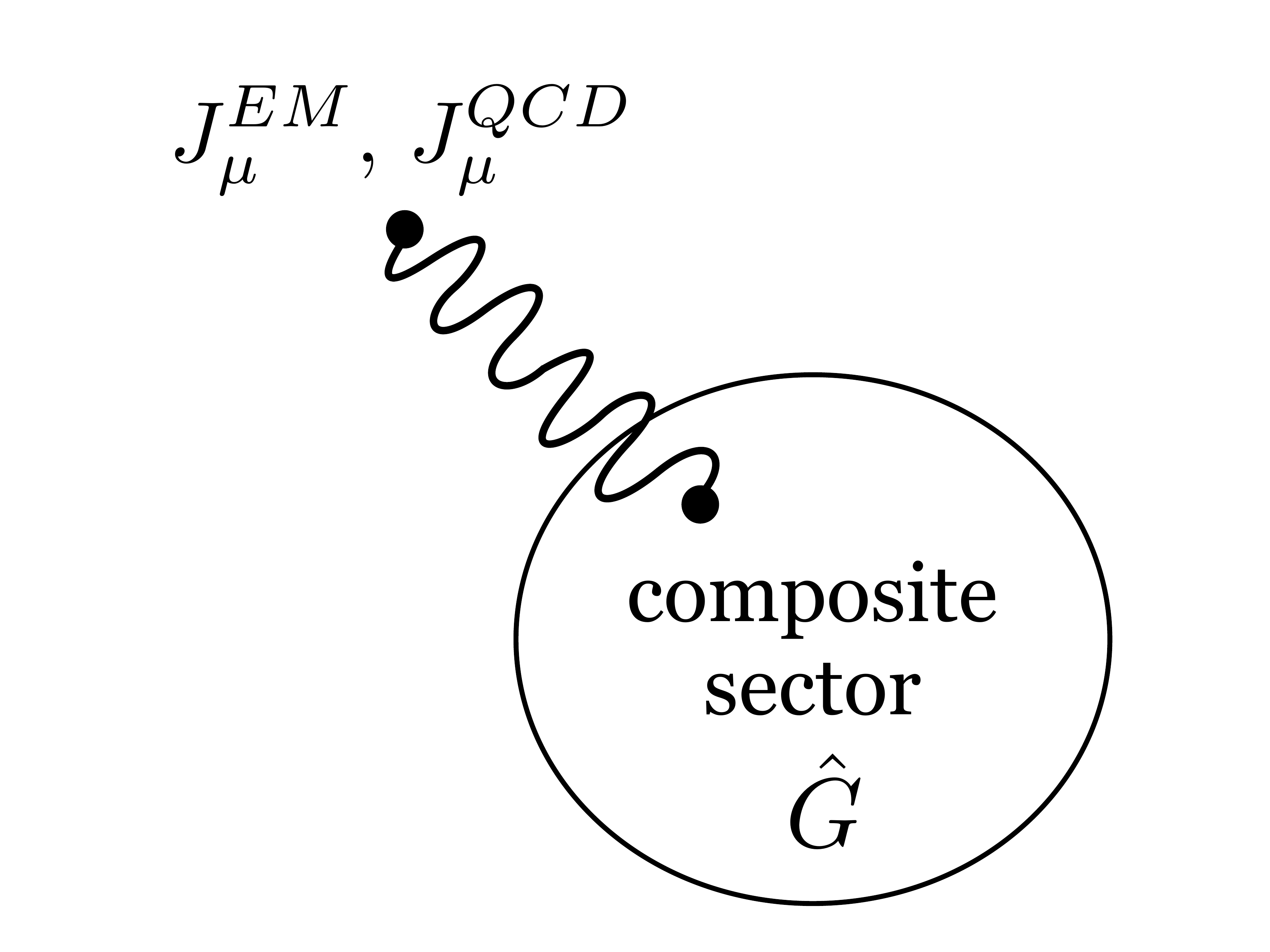} \label{AtoCFT}
\caption{Sourcing an operator to the composite sector through external sources.}
\end{figure}

Within the same dual picture one can describe bulk gauge fields. If the compactification preserves gauge invariance at the level of zero modes, that situation corresponds to a global symmetry in the 4D sector which has been weakly gauged through adding external sources $J_{\mu}$ to the strong sector, switching on some new operators ${\cal O}^{\mu}$,
\bea
{\cal L}_{comp} \supset g_{comp} \, J_{\mu} {\cal O}^{\mu} \ ,
\eea
  as we schematically represent in Fig.~\ref{AtoCFT}.

 The key question for holography is whether there is always a metric $w(z)$ and field configuration, no matter how complicated it is, which is dual to our 4D target theory. And this paper showed an instance where this is not the case. Indeed, $\hat{G}$ (or the preon quarks or bosons composing $\hat{G}$) could have no electric or color charges, leading to a ratio $R\in(0,\infty)$. In Fig.~\ref{AtoCFT}, this corresponds to setting one, or both, of the sources to zero. Note, though, that this study is largely based on s-channel resonance by gluon-initiated processes, so that one would compare the effect of $G$ with a $\hat{G}$ with colored constituents. The Regge gluon \cite{Perelstein:2009qi,Perelstein:2011ez}, a spin-2 excitation of the Standard Model gluon in warped extra dimension, is an explicit example of this scenario, There, the Regge gluon couples to the Standard Model gluons but it does not interact with photons at tree level.

\section{Conclusions}

If new strong interactions lurk near the electroweak scale, one expects a rich variety of new resonances, both mesonic and baryonic. New strong interactions may not deconfine before energies well beyond LHC reach. Instead of finding evidence of form factor interactions or production of the preons, only a suppressed compositeness behavior would be accessible.

The question then becomes one of identifying the new sector without really accessing its perturbative description. But composite fermions or vector bosons can mimic new matter generations and new spontaneously-broken gauge symmetries respectively, hindering their unambiguous identification as composite or elementary.

As for spin-two resonances, only one framework of new physics is able to mimic them: Kaluza-Klein gravitons. In this paper, we revised the claim that spin-two resonances are a smoking gun for extra dimensions, and were able distinguish between the two scenarios, i.e. spin-2 resonances vs. KK gravitons.

Distinguishing between the KK graviton and the impostor turns out to be harder than first expected. Although gravity couples to fields in a very constrained manner, after compactification, there is quite a lot of model dependence in the coupling strength to the operators in the stress tensor. Still, one could have expected that the impostor would couple to different operators than the KK graviton, hence leading to a clear signature of new strong dynamics. But Lorentz invariance and the SM gauge, flavor and CP symmetries are so restrictive that the impostor ends up coupling to the same structures as the massive graviton.

Nevertheless, we found a robust prediction for the decays of the KK graviton, and propose this measurement as a way of distinguishing between new extra dimensions and  new strong interactions. 

\section*{Acknowledgements}
V. Sanz thanks L. Randall for useful comments.
This work is partially supported by funding from NSERC of Canada.


\end{document}